\documentclass{article}
\usepackage[latin9]{inputenc}
\usepackage{url}
\usepackage{multirow}
\usepackage{amstext}
\usepackage[authoryear]{natbib}

\makeatletter

\providecommand{\tabularnewline}{\\}

\newcommand{\lyxaddress}[1]{
	\par {\raggedright #1
	\vspace{1.4em}
	\noindent\par}
}

\makeatother

\begin{document}
\title{New Equivalence Tests for Hardy-Weinberg Equilibrium and Multiple
Alleles}
\author{Vladimir Ostrovski\thanks{vladimir.ostrovski77@googlemail.com}}
\maketitle

\lyxaddress{ERGO Group AG, ERGO-Platz 1, 40198 Düsseldorf, Germany}
\begin{abstract}
We consider testing equivalence to Hardy-Weinberg Equilibrium in case
of multiple alleles. Two different test statistics are proposed for
this test problem. The asymptotic distribution of the test statistics
is derived. The corresponding tests can be carried out using asymptotic
approximation. Alternatively, the variance of the test statistics
can be estimated by the bootstrap method. The proposed tests are applied
to three real data sets. The finite sample performance of the tests
is studied by simulations, which are inspired by the real data sets.
\end{abstract}

\section{Introduction}

Hardy-Weinberg Equilibrium (HWE) plays an important role in the field
of the population genetics and related scientific domains. HWE is
a common assumption in many areas of research so that assessing the
compatibility of observed genotype frequencies with HWE is a basic
step of a complete statistical analysis. There are two main approaches
to this undertaking: goodness of fit tests and equivalence tests.

A vast amount of literature exists on the goodness of fit tests for
HWE, which includes application of the asymptotic $\chi^{2}$ and
likelihood ratio tests. The specific exact goodness of fit tests for
HWE are developed in \citet{Levene1949}, \citet{Haldane_1954}, \citet{Chapco1976},
\citet{LouisDempster1987} and \citet{GuoThompson1992} among others.
The null hypothesis of all these tests is that the underlying population
is exactly in HWE. Hence, the goodness of fit tests are tailored to
establish lack of compatibility with HWE.

The equivalence tests are appropriate to establish sufficiently good
agreement of the observed genotype frequencies with HWE. The exact
and approximate equivalence tests for the biallelic case are developed
recently in \citet{Wellek2004}, \citet{Wellek2010} and \citet{WelleKGoddardZiegler2010}.
To our best knowledge, there are not any equivalence tests for HWE
and multiple alleles. Two different equivalence tests are developed
in this paper for the case of multiple alleles. The tests can be carried
out using the asymptotic approximation or bootstrap method.

A distribution of diploid genotypes at a $k$-allele locus can be
represented as a lower triangular matrix $p$, where $p\left(i,j\right)$
is the probability of the genotypes with alleles $i$ and $j$. Let
$a\left(p\right)$ denote the allele distribution under $p$. The
probability of the allele $i$ under $p$ can be calculated as $a\left(p,i\right)=\frac{1}{2}\sum_{j=1}^{k}\left(p\left(i,j\right)+p\left(j,i\right)\right)$.
If the population is in HWE, then the genotype distribution fulfills
the conditions $p(i,j)=2a(p,i)a(p,j)$ for $i<j$ and $p(i,i)=a(p,i)^{2}$.
Let $e\left(p\right)$ denote the genotype distribution under the
assumption of HWE, which is implied by the allele distribution $a\left(p\right)$.

Euclidean distance $l_{2}\left(p,e(p)\right)$ can be considered a
conditional distance between the genotype distributions $p$ and $e\left(p\right)$
under the joint allele distribution $a\left(p\right)$. The equivalence
test problem is then defined by 
\begin{eqnarray}
H_{0}=\left\{ l_{2}\left(p,e\left(p\right)\right)\geq\varepsilon\right\}  & \text{and} & H_{1}=\left\{ l_{2}\left(p,e\left(p\right)\right)<\varepsilon\right\} \label{condTestProblem}
\end{eqnarray}
where $\varepsilon$ is a tolerance parameter.

Let $\mathcal{M}$ denote the family of all possible genotype distributions
at HWE. The minimum distance between $p$ and $\mathcal{M}$ is defined
by $d\left(p,\mathcal{M}\right)=\min_{q\in\mathcal{M}}l_{2}\left(p,q\right)$.
The corresponding equivalence test problem is given by

\begin{eqnarray}
H_{0}=\left\{ d\left(p,\mathcal{M}\right)\geq\varepsilon\right\}  & \text{and} & H_{1}=\left\{ d\left(p,\mathcal{M}\right)<\varepsilon\right\} \label{minDstTestPr}
\end{eqnarray}
We observe the genotype frequencies $p_{n}$ of the sample size $n$.
The natural test statistic for (\ref{condTestProblem}) is 
\[
T_{c}\left(p_{n}\right)=\sqrt{n}\left(l_{2}^{2}\left(p_{n},e\left(p_{n}\right)\right)-\varepsilon^{2}\right),
\]
which can be easily computed. The appropriate test statistic for (\ref{minDstTestPr})
is 
\[
T_{m}\left(p_{n}\right)=\sqrt{n}\left(d^{2}\left(p_{n},\mathcal{M}\right)-\varepsilon^{2}\right),
\]
which requires optimization for the calculation of $d\left(p_{n},\mathcal{M}\right)$.
The test statistic $T_{c}\left(p_{n}\right)$ can be considered a
numerically efficient approximation to $T_{m}\left(p_{n}\right)$
because of $l_{2}\left(p_{n},e\left(p_{n}\right)\right)\geq d\left(p_{n},\mathcal{M}\right)$.
The subscript $*$ will be used instead of $c$ and $m$ in the reminder
of the paper, if statements are appropriate for both cases.

\section{Equivalence tests}

In this section, we derive the asymptotic distributions of the test
statistics $T_{c}\left(p_{n}\right)$ and $T_{m}\left(p_{n}\right)$.
We provide also an algorithm for the asymptotic and bootstrap based
tests.

Let $v$ be the usual bijective mapping of the matrix $p$ to the
vector $\left(p(1,1),p\left(1,2\right),\ldots,p\left(k,k\right)\right)$.
Let $\mathring{d_{c}}$ denote the derivative of the function $q\mapsto l_{2}^{2}\left(v^{-1}\left(q\right),e\left(v^{-1}\left(q\right)\right)\right)$,
where $q$ is a vector of length $k^{2}$. The derivative $\mathring{d_{c}}$
can be derived using the chain rule. Let $p_{0}\in H_{0}$ fulfill
the boundary condition $l_{2}\left(p_{0},e\left(p_{0}\right)\right)=\varepsilon$
and let $q_{0}=v\left(p_{0}\right)$. Then the asymptotic distribution
of $T_{c}\left(p_{n}\right)$ under $p_{0}$ is Gaussian with mean
zero and variance $\sigma_{c}^{2}\left(p_{0}\right)=\mathring{d}_{c}\left(q_{0}\right)\Sigma\left(q_{0}\right)\mathring{d}_{c}\left(q_{0}\right)^{t}$,
where $\Sigma\left(q_{0}\right)=D_{q}-qq^{t}$ is a covariance matrix
and $D_{q}$ is a square diagonal matrix, whose diagonal entries are
elements of $q$. The proof of the statement can be found in \citet{Ostrovski2019}.

The test statistic $T_{m}\left(p_{n}\right)$ converges weakly under
the assumption, that there exists a continuous function $h$ on an
open neighborhood of $p_{0}$ such that $h\left(p\right)\in\mathcal{M}$
and $d\left(p,\mathcal{M}\right)=l_{2}\left(p,h\left(p\right)\right)$.
The existence of a continuous minimizer $h$ is also an important
requirement for the numerical computation of $d\left(p,\mathcal{M}\right)$.
We assume the existence of a continuous minimizer $h$ on an open
neighborhood of $p_{0}$ for the reminder of the paper. Let $\mathring{d_{m}}$
denote the derivative of the function $q\mapsto l_{2}^{2}\left(v^{-1}\left(q\right),h\left(p_{0}\right)\right)$.
Then the asymptotic distribution of $T_{m}\left(p_{n}\right)$ under
$p_{0}$ is Gaussian with mean zero and variance $\sigma_{m}^{2}\left(p_{0}\right)=\mathring{d}_{m}\left(q_{0}\right)\Sigma\left(q_{0}\right)\mathring{d}_{m}\left(q_{0}\right)^{t}$,
see \citet{Ostrovski2018} for details.

The asymptotic variance $\sigma_{*}^{2}\left(p_{0}\right)$ is unknown
and can be estimated by $\sigma_{*}^{2}\left(p_{n}\right)$. The asymptotic
test can be carried out as follows: 
\begin{enumerate}
\item Given are the genotype frequencies $p_{n}$, the tolerance parameter
$\varepsilon$ and the significance level $\alpha$. 
\item Compute the tests statistic $T_{*}\left(p_{n}\right)$. 
\item Estimate the asymptotic variance by $\sigma_{*}^{2}\left(p_{n}\right)$. 
\item Reject $H_{0}$ if $T_{*}\left(p_{n}\right)\leq c_{\alpha}\sigma_{*}\left(p_{n}\right)$,
where $c_{\alpha}$ is the lower $\alpha$-quantile of the normal
distribution. 
\end{enumerate}
The minimum tolerance parameter $\varepsilon$, for which the asymptotic
test can reject $H_{0}$, can be computed as $\sqrt{l_{2}^{2}\left(p_{n},e\left(p_{n}\right)\right)-n^{-\frac{1}{2}}c_{\alpha}\sigma_{c}\left(p_{n}\right)}$
or $\sqrt{d\left(p_{n},\mathcal{M}\right)-n^{-\frac{1}{2}}c_{\alpha}\sigma_{m}\left(p_{n}\right)}$
correspondingly.

In order to improve the finite sample performance of the proposed
tests, the bootstrap method is applied to estimate the variance of
$T_{*}\left(p_{n}\right)$, see \citet{EfronTibshirani1993}, Section
6 for details. The estimator $\sigma_{*}\left(p_{n}\right)$ is then
replaced by the bootstrap estimator of the variance. Otherwise, everything
stays the same.

\section{Simulation study}

The proposed tests are implemented in R and are freely available on
GitHub under \url{https://github.com/TestingEquivalence/HardyWeinbergEquilibriumR}.
All simulations are performed in R-Studio on a usual scientific workstation.

\subsection{Real data sets}\label{subsec:Real-data-sets}

The equivalence tests are applied to the following data sets, which
are already analyzed in the literature on the goodness of fit tests:
1. from rheumatoid arthritis study, \citet{WordsworthAtEl1992}; 2.
from the documentation included with the GENEPOP software package,
\citet{Rousset2008}; 3. genotype frequency data at Rhesus locus,
\citet{CavalliBodmer1971}. The genotype distributions of the data
sets are given in Table \ref{data_sets}.

The minimum tolerance parameters $\varepsilon$, for which the tests
can reject the corresponding $H_{0}$, are displayed in Table \ref{test_res}.
The distances $d\left(p_{n},\mathcal{M}\right)$ and $l_{2}\left(p_{n},e\left(p_{n}\right)\right)$
are close to each other in all cases so that $l_{2}\left(p_{n},e\left(p_{n}\right)\right)$
provides a good approximation to $d\left(p_{n},\mathcal{M}\right)$.
The test results are also similar for $T_{c}$ and $T_{m}$. The bootstrap
tests are slightly more conservative than the asymptotic tests in
all cases.

It could not be shown that data sets 1 and 2 are close to HWE. All
goodness of fit tests in \citet{Engels2009} reject also the null
hypothesis of HWE for data set 2 at the nominal level 0.05. Data set
3 is very close to HWE. This observation corresponds to the results
of the goodness of fit tests in \citet{GuoThompson1992} and \citet{Engels2009}.

\begin{table}
\begin{tabular}{|c|c|c|c|cc|cc|}
\hline 
\multirow{1}{*}{data set} & \multirow{1}{*}{$n$} & \multirow{1}{*}{$l_{2}\left(p_{n},e\left(p_{n}\right)\right)$} & \multirow{1}{*}{$d\left(p_{n},\mathcal{M}\right)$} & \multicolumn{2}{c|}{$T_{c}$} & \multicolumn{2}{c|}{$T_{m}$}\tabularnewline
 &  &  &  & A  & B  & A  & B\tabularnewline
\hline 
1  & 230  & 0.102  & 0.101  & 0.130  & 0.134  & 0.130  & 0.132\tabularnewline
2  & 229  & 0.126  & 0.118  & 0.159  & 0.164  & 0.149  & 0.153\tabularnewline
3  & 8295  & 0.013  & 0.013  & 0.017  & 0.019  & 0.018  & 0.018\tabularnewline
\hline 
\end{tabular}

\caption{Minimum tolerance parameter $\varepsilon$, for which $H_{0}$ can
be rejected at the nominal level $0.05$. A stands for the asymptotic
test and B stands for the bootstrap test. }\label{test_res}
\end{table}

\subsection{Test power}

In this subsection we study the test power at HWE. We restrict yourself
to the genotype distributions at HWE, which are implied by the real
data sets from Subsection \ref{subsec:Real-data-sets}, because the
family $\mathcal{M}$ is very large. In order to shed some light on
the appropriate values of the tolerance parameter $\varepsilon$,
the test power is computed for different values of $\varepsilon$,
see Table \ref{test_power}. The value of $\varepsilon$ may be considered
appropriate if the test power is approximately 0.9. Hence, the appropriate
value of $\varepsilon$ is 0.1 for data set 1, 0.1 for data set 2
and 0.018 for data set 3.

\begin{table}
\begin{tabular}{|c|cccc|cccc|cccc|}
\hline 
 & \multicolumn{4}{c|}{data set 1} & \multicolumn{4}{c|}{data set 2} & \multicolumn{4}{c|}{data set 3}\tabularnewline
\hline 
$\varepsilon$  & 0.07  & 0.08  & 0.09  & 0.10  & 0.07  & 0.08  & 0.09  & 0.10  & 0.012  & 0.014  & 0.016  & 0.018\tabularnewline
\hline 
$T_{c}$, A  & 0.56  & 0.75  & 0.87  & 0.95  & 0.53  & 0.74  & 0.87  & 0.94  & 0.67  & 0.82  & 0.91  & 0.96\tabularnewline
$T_{c}$, B  & 0.40  & 0.63  & 0.79  & 0.90  & 0.40  & 0.62  & 0.80  & 0.90  & 0.50  & 0.71  & 0.83  & 0.91\tabularnewline
\hline 
$T_{m}$, A  & 0.54  & 0.74  & 0.87  & 0.94  & 0.58  & 0.78  & 0.89  & 0.96  & 0.61  & 0.77  & 0.88  & 0.95\tabularnewline
$T_{m}$, B  & 0.49  & 0.70  & 0.85  & 0.93  & 0.53  & 0.74  & 0.86  & 0.94  & 0.58  & 0.75  & 0.86  & 0.94\tabularnewline
\hline 
\end{tabular}

\caption{Simulated rejection probability of the equivalence tests at the nominal
level $0.05$. The rejection probability is simulated for different
values of the tolerance parameter $\varepsilon$ at the HWE distributions
$e\left(p_{n}\right)$, which are implied by data sets 1, 2 and 3.
The sample size equals the size of the corresponding data set. The
number of replications is 1000 for each experiment. A stands for the
asymptotic test and B stands for the bootstrap test. }\label{test_power}
\end{table}

The observed genotype frequencies $p_{n}$ are subjected to the sampling
error. It is important for the test efficiency that the sampling error
has small influence on the test power at HWE. The test power is computed
at 100 random genotype frequencies, where the corresponding random
samples of size $n$ are drawn from the implied genotype distribution
$e\left(p_{n}\right)$. The simulation results are summarized in Table
\ref{test_power_sensi}. The power of all considered tests varies
little from point to point. Hence, the impact of the sampling error
on the test power at HWE is very small.

\begin{table}
\begin{tabular}{|c|cccc|cccc|cccc|}
\hline 
 & \multicolumn{4}{c|}{data set 1, $\varepsilon=0.1$} & \multicolumn{4}{c|}{data set 2, $\varepsilon=0.1$} & \multicolumn{4}{c|}{data set 3, $\varepsilon=0.018$}\tabularnewline
\hline 
 & min  & max  & mean  & dev  & min  & max  & mean  & dev  & min  & max  & mean  & dev\tabularnewline
\hline 
$T_{c}$, A  & 0.93  & 0.97  & 0.95  & 0.008  & 0.93  & 0.97  & 0.94  & 0.008  & 0.94  & 0.98  & 0.97  & 0.007\tabularnewline
$T_{c}$, B  & 0.87  & 0.92  & 0.90  & 0.010  & 0.88  & 0.94  & 0.90  & 0.012  & 0.89  & 0.94  & 0.91  & 0.010\tabularnewline
\hline 
$T_{m}$, A  & 0.92  & 0.96  & 0.95  & 0.009  & 0.93  & 0.99  & 0.96  & 0.012  & 0.92  & 0.97  & 0.95  & 0.007\tabularnewline
$T_{m}$, B  & 0.90  & 0.95  & 0.93  & 0.010  & 0.91  & 0.99  & 0.95  & 0.015  & 0.91  & 0.96  & 0.94  & 0.009\tabularnewline
\hline 
\end{tabular}

\caption{Summary of the simulated rejection probabilities at the nominal level
$0.05$. The rejection probabilities are simulated at the 100 random
samples from the HWE distributions $e\left(p_{n}\right)$, which are
implied by data sets 1, 2 and 3. The sample size equals the size of
the corresponding data set. The number of replications is 1000 for
each experiment. A stands for the asymptotic test and B stands for
the bootstrap test.}\label{test_power_sensi}
\end{table}

\subsection{Type I error}

We study the type I error rates of the proposed tests in this subsection.
The boundary of $H_{0}$ is so complex that it is very difficult to
find boundary points, which have the largest rejection probability.
We consider therefore randomly selected boundary points of $H_{0}$,
which are based on the three real data sets from Subsection \ref{subsec:Real-data-sets}.
The boundary points are generated using the following algorithm: 
\begin{enumerate}
\item Given are $p_{n}$ and $\varepsilon$. 
\item Draw a sample of size $n$ from $p_{n}$ and compute the sample genotype
frequency $\widetilde{p}_{n}$. 
\item If $T_{*}\left(\widetilde{p}_{n}\right)<0$ then reject $\widetilde{p}_{n}$
and repeat step 2. Otherwise accept $\widetilde{p}_{n}$. 
\item Consider the linear combination $a\widetilde{p}_{n}+\left(1-a\right)e\left(p_{n}\right)$
for $a\in\left[0,1\right]$. Find $a_{n}\in\left[0,1\right]$ such
that $T_{*}\left(a_{n}\widetilde{p}_{n}+\left(1-a_{n}\right)e\left(p_{n}\right)\right)=0$.
The value of $a_{n}$ can be found using any line search method. 
\item Return $a_{n}\widetilde{p}_{n}+\left(1-a_{n}\right)e\left(p_{n}\right)$,
which is a random boundary point of $H_{0}$. 
\end{enumerate}
The tolerance parameter $\varepsilon$ is close to $l_{2}\left(p_{n},e\left(p_{n}\right)\right)$
for each data set under consideration so that $a_{n}$ is usually
not far from 1. The corresponding random boundary point is then close
to $p_{n}$. Hence, we explore the boundary of $H_{0}$ in the neighborhood
of the given data set. The test power at 100 random boundary points
is summarized in Table \ref{test_power_boundary}. The test power
varies considerable from point to point. The asymptotic test based
on $T_{c}$ is not conservative for all three data sets. The asymptotic
test based on $T_{m}$ shows some anti-conservative tendencies for
data sets 2 and 3. The bootstrap test based on $T_{c}$ is conservative
for all three data sets. The bootstrap test based on $T_{m}$ shows
slight non conservative tendencies. If more conservative tests are
required then the nominal level may be halved or the tolerance parameter
may be reduced.

We recommend to perform all proposed tests in any case and compare
the results. Additionally, the rejection probabilities at the close
random boundary points may be studied by means of simulation.

\begin{table}
\begin{tabular}{|c|cccc|cccc|cccc|}
\hline 
 & \multicolumn{4}{c|}{data set 1, $\varepsilon=0.1$} & \multicolumn{4}{c|}{data set 2, $\varepsilon=0.1$} & \multicolumn{4}{c|}{data set 3, $\varepsilon=0.018$}\tabularnewline
\hline 
 & min  & max  & mean  & dev  & min  & max  & mean  & dev  & min  & max  & mean  & dev\tabularnewline
\hline 
$T_{c}$, A  & 0.017  & 0.060  & 0.035  & 0.009  & 0.023  & 0.065  & 0.044  & 0.009  & 0.051  & 0.114  & 0.090  & 0.011\tabularnewline
$T_{c}$, B  & 0.005  & 0.036  & 0.018  & 0.006  & 0.013  & 0.048  & 0.029  & 0.006  & 0.012  & 0.052  & 0.033  & 0.008\tabularnewline
\hline 
$T_{m}$, A  & 0.019  & 0.049  & 0.031  & 0.006  & 0.025  & 0.077  & 0.046  & 0.009  & 0.034  & 0.075  & 0.051  & 0.008\tabularnewline
$T_{m}$, B  & 0.011  & 0.042  & 0.022  & 0.006  & 0.016  & 0.064  & 0.034  & 0.008  & 0.025  & 0.064  & 0.038  & 0.008\tabularnewline
\hline 
\end{tabular}

\caption{Summary of the simulated rejection probabilities at the nominal level
$0.05$. The rejection probabilities are simulated at the 100 randomly
selected boundary points of $H_{0}$. The sample size equals the size
of the corresponding data set. The number of replications is 1000
for each experiment. A stands for the asymptotic test and B stands
for the bootstrap test.}\label{test_power_boundary}
\end{table}

\begin{table}
\begin{minipage}[t][1\totalheight][s]{0.25\columnwidth}%
\begin{tabular}{|cccc|}
\hline 
5  &  &  & \tabularnewline
40  & 12  &  & \tabularnewline
6  & 32  & 2  & \tabularnewline
30  & 55  & 15  & 33\tabularnewline
\hline 
\end{tabular}

1)%
\end{minipage}%
\begin{minipage}[t]{0.25\columnwidth}%
\begin{tabular}{|cccc|}
\hline 
2  &  &  & \tabularnewline
12  & 24  &  & \tabularnewline
30  & 34  & 54  & \tabularnewline
22  & 21  & 20  & 10\tabularnewline
\hline 
\end{tabular}

2)%
\end{minipage}

\begin{minipage}[t]{0.45\columnwidth}%
\begin{tabular}{|ccccccccc|}
\hline 
1236  &  &  &  &  &  &  &  & \tabularnewline
120  & 3  &  &  &  &  &  &  & \tabularnewline
18  & 0  & 0  &  &  &  &  &  & \tabularnewline
982  & 55  & 7  & 249  &  &  &  &  & \tabularnewline
32  & 1  & 0  & 12  & 0  &  &  &  & \tabularnewline
2582  & 132  & 20  & 1162  & 29  & 1312  &  &  & \tabularnewline
6  & 0  & 0  & 4  & 0  & 4  & 0  &  & \tabularnewline
2  & 0  & 0  & 0  & 0  & 0  & 0  & 0  & \tabularnewline
115  & 5  & 2  & 53  & 1  & 149  & 0  & 0  & 4\tabularnewline
\hline 
\end{tabular}

3)%
\end{minipage}\caption{The data sets: 1) from rheumatoid arthritis study, \citet{WordsworthAtEl1992};
2) from the documentation included with the GENEPOP software package,
\citet{Rousset2008}; 3) genotype frequency data at Rhesus locus,
\citet{CavalliBodmer1971}.}\label{data_sets}
\end{table}

\bibliographystyle{plainnat}
\bibliography{VO_literature}

\end{document}